\documentclass[preprint,journal]{vgtc}       %

\ifpdf%
  \pdfoutput=1\relax                   %
  \usepackage{graphicx}                %
  \DeclareGraphicsExtensions{.pdf,.png,.jpg,.jpeg} %
\else%
  \usepackage{graphicx}                %
  \DeclareGraphicsExtensions{.eps}     %
\fi%

\graphicspath{{figures/}{pictures/}{images/}{./}} %

\usepackage{microtype}                 %
\PassOptionsToPackage{warn}{textcomp}  %
\usepackage{textcomp}                  %
\usepackage{mathptmx}                  %
\usepackage{times}                     %
\usepackage{cite}                      %
\usepackage{tabu}                      %
\usepackage{booktabs}                  %
\usepackage[dvipsnames, table]{xcolor}
\usepackage{amssymb}
\usepackage{soul}
\usepackage{enumitem}

\preprinttext{To appear in IEEE Transactions on Visualization and Computer Graphics, January 2021.}

\onlineid{1147}

\vgtccategory{Research}
\vgtcpapertype{algorithm/technique}

\title{Selection-Bias-Corrected Visualization via Dynamic Reweighting}

\author{David Borland, Jonathan Zhang, Smiti Kaul, and David Gotz}
\authorfooter{
\item
 David Borland is with RENCI at the University of North Carolina at Chapel Hill. E-mail: borland@renci.org.
\item
 Jonathan Zhang is with the Dept. of Biostatistics at the University of North Carolina at Chapel Hill. E-mail: jzhang42@live.unc.edu.
\item
 Smiti Kaul is with the Dept. of Computer Science at the University of North Carolina at Chapel Hill. E-mail: smiti@live.unc.edu.
\item
 David Gotz is with the School of Information and Library Science at the University of North Carolina at Chapel Hill. E-mail: gotz@unc.edu.
}

\shortauthortitle{Borland \MakeLowercase{\textit{et al.}}: Selection-Bias-Corrected Visualization via Dynamic Reweighting}

\abstract{The collection and visual analysis of large-scale data from complex systems, such as electronic health records or clickstream data, has become increasingly common across a wide range of industries.  This type of retrospective visual analysis, however, is prone to a variety of selection bias effects, especially for high-dimensional data where only a subset of dimensions is visualized at any given time. The risk of selection bias is even higher when analysts dynamically apply filters or perform grouping operations during ad hoc analyses. These bias effects threatens the validity and generalizability of insights discovered during visual analysis as the basis for decision making. Past work has focused on \emph{bias transparency}, helping users understand when selection bias may have occurred. However, countering the effects of selection bias via \emph{bias mitigation} is typically left for the user to accomplish as a separate process. Dynamic reweighting (DR) is a novel computational approach to selection bias mitigation that helps users craft bias-corrected visualizations.  This paper describes the DR workflow, introduces key DR visualization designs, and presents statistical methods that support the DR process.  Use cases from the medical domain, as well as findings from domain expert user interviews, are also reported.
}


\CCScatlist{ %
 \CCScat{K.6.1}{Management of Computing and Information Systems}%
{Project and People Management}{Life Cycle};
 \CCScat{K.7.m}{The Computing Profession}{Miscellaneous}{Ethics}
}

\teaser{
  \centering
  \includegraphics[width=\linewidth]{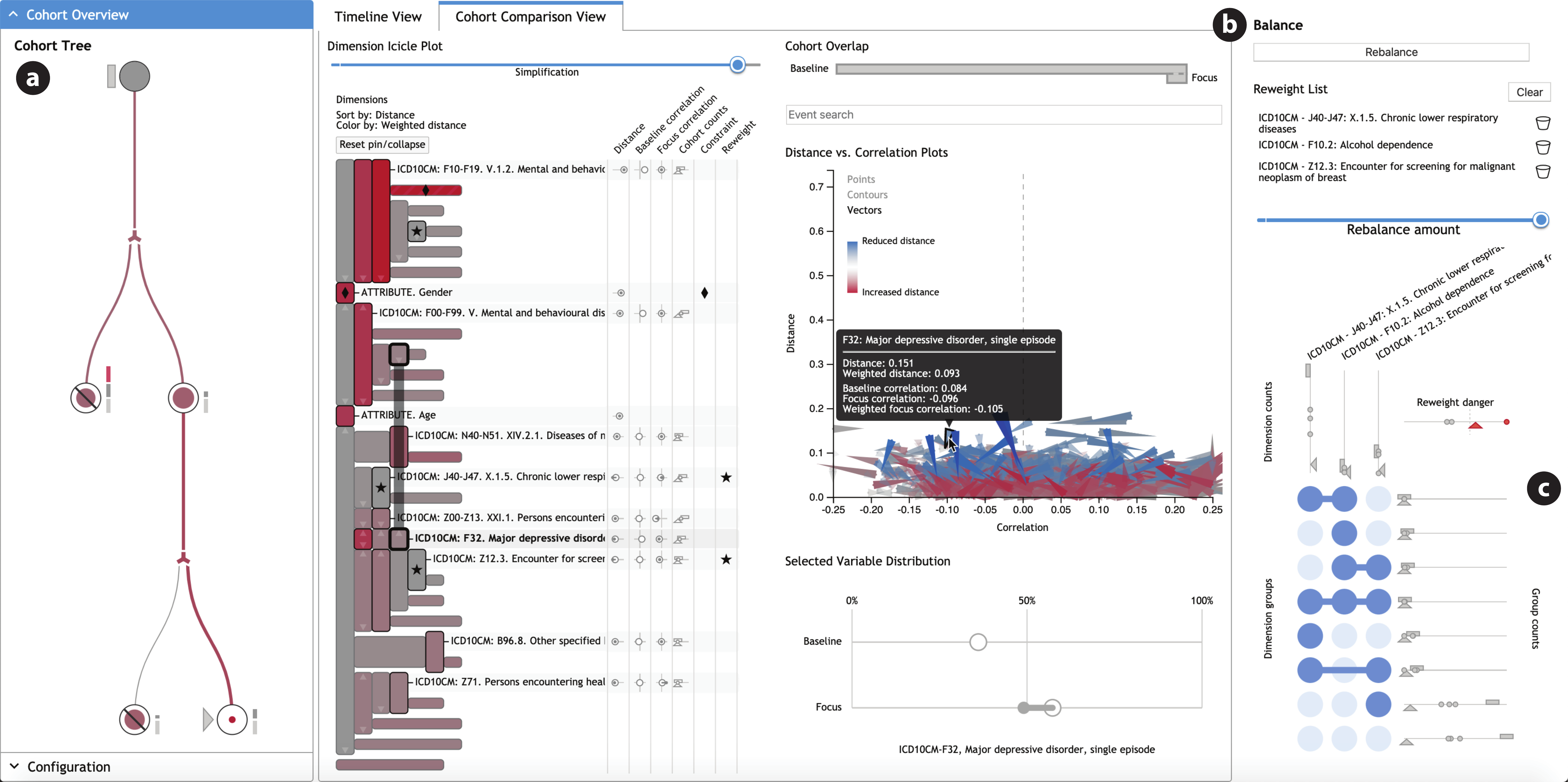}
  \caption{
  Coordinated visualization panels supporting the dynamic reweighting workflow. The visual designs help users discover dimensions with high selection bias, select dimensions for correction, and assess both the quality and impact of a reweighting solution.}
	\label{fig:teaser}
}

 \nocopyrightspace

\vgtcinsertpkg

\begin{document}

\firstsection{Introduction}

\maketitle

Large-scale data collection and analysis has become common across a wide range of domains.
As reflected by the quickly growing and evolving visualization industry \cite{behrisch_commercial_2019}, visual analytics is a key enabler for many of these applications, in part due to the promise of providing analysts and decision makers with intuitive tools to quickly explore and reason about data \cite{pak_chung_wong_visual_2004}.  Ideally, these tools enable analysts to interpret rich
datasets collected from complex systems
and make nuanced inferences that reflect the complexity of real-world situations.  

However, retrospective visual analyses of data collected ``in the wild''---in contrast to
data from randomized controlled trials---are subject to a variety of selection bias effects \cite{hernan_structural_2004} that limit the ability to generalize from the insights gained through the visualizations.

In fields such as political election polling, selection bias often arises because the sample---those who answer the polls---is often different from the population of interest---those who are going to vote---across various demographics (e.g. age, race, and socioeconomic status). Responses from subgroups of the sample defined by these demographics are often weighted to achieve more accurate results with respect to the larger population. The risk of selection bias is especially acute for interactive visualizations in which users can quickly apply filters or grouping operations to define data subsets during exploratory analysis \cite{borland_contextual_2018}. This problem is compounded when working with high-dimensional data where only a subset of variables are visualized at any given time. Such situations can result in hidden bias effects that go unnoticed by the user.
    
To address this and similar bias-related challenges, a number of recent research efforts have explored new methods for \emph{bias transparency}.  These techniques aim to quantify and visually communicate the threat of bias as it emerges during visual analysis \cite{wall_warning_2017,borland_contextual_2018}. This passive approach can alert users to what would otherwise be invisible threats to the validity of an analysis, with the hope that users will be able to adjust their analysis activity or interpretation of the data in response.  

This strategy has been shown to be effective in some cases, including for selection bias  \cite{gotz_adaptive_2016}.  However, while bias transparency methods can help users understand the limitations of their analyses, they fall short in helping users overcome those limitations. 
For example, consider a recent visual analytics system for cohort selection using medical data \cite{borland_selection_2020}. The system provides bias transparency, but users have only three options if a large bias is revealed:
(1) experiment with changes to their inclusion/exclusion criteria with the goal of reducing selection bias (often difficult if not impossible depending on the makeup of the original dataset), (2) initiate a new data collection effort to obtain a better initial sample of patient data (often cost prohibitive if not impossible), or (3) abort the analysis entirely.

This paper presents \emph{dynamic reweighting} (DR), a new approach for \emph{bias mitigation} which goes beyond solely communicating selection bias effects to enable selection bias correction during interactive visual analysis of high-dimensional data.  DR is a workflow---supported by a set of interactive visualization designs---that produces weighted data samples and the resultant weighted aggregate statistics. DR enables the creation of bias-corrected visualizations and provides analysts with the tools required to guide and assess the bias correction process. 
More specifically, the research contributions presented in this paper include:

\begin{itemize}
    \item {\bf Dynamic Reweighting (DR) Workflow:} A user-driven workflow for guiding the creation of bias-corrected visualizations.
\vspace{-0.15cm}
    \item {\bf DR Visualizations:} New visualization designs for key steps in the DR workflow, including (1) selecting data dimensions for bias correction, (2) assessing the quality of a DR solution, and (3) exploring the impact of bias correction on the visualized results.
\vspace{-0.15cm}    
    \item {\bf DR Reweighting Methods:} A set of statistical methods that support DR, including class segmentation, weight calculation, and computation of a ``danger score'' quality assessment measure.
\end{itemize}

This paper describes the contributions outlined above and their implementation within the \textit{Cadence} visual analytics platform for cohort selection, presents an example use case for bias correction, and reports feedback obtained from interviews with medical domain experts.

\section{Related Work}

The research presented in this paper builds upon previous work relevant to bias mitigation, including applied statistical research in survey science and epidemiology.
The application used for the case study and interviews leverages
cohort selection and event sequence visual analytics techniques, and the DR visualizations
are grounded in prior research in hierarchy and set visualization, among other areas.

\subsection{Selection Bias as a Threat to Validity} 

The literature on research study design has long focused on the importance of accounting for common threats to validity when gathering or analyzing data~\cite{brewer_research_2000}. Selection bias (a.k.a. sample bias) is widely recognized as a key challenge that threatens both internal and external validity~\cite{hernan_structural_2004}. It can manifest either in the original sampling of a dataset
or during analysis by conditioning the inclusion of certain variables during filtering or grouping operations.

To address the importance of selection bias on study validity, various methods to minimize bias have been proposed for data collection and management (e.g., randomized controlled trials~\cite{jadad_randomised_1998}). However, many visual analysis tools are applied to data collected through less careful mechanisms. Moreover, given the ad hoc nature of exploratory analysis, careful sampling of the visualized data is not typically practical.

In more traditional statistical settings, techniques such as propensity score matching (PSM)~\cite{caliendo_practical_2008} have been used to apply adjustments to poorly sampled data. However, because PSM aims to adjust all variable distributions simultaneously, it has been shown to often increase imbalance and bias \cite{king_why_2019}. 
The DR approach is more closely related to the sample weighting methods used in survey statistics~\cite{gary_adjusting_2007}, in which adjustments are applied to groups of samples based on combinations of a small number of variables. Traditionally, only a few predefined variables are used to determine the groups used for calculating a static set of weights for a given analysis.  However, challenges of high-dimensionality, along with the ad hoc nature of exploratory visual analysis, render statistical adjustment even more challenging~\cite{gelman_struggles_2007}, thus motivating the new methods proposed in this paper.

\subsection{Selection Bias in Visual Analytics}

Bias, including techniques for both transparency~\cite{borland_contextual_2018,wall_warning_2017,gotz_adaptive_2016} and mitigation~\cite{wall_toward_2019}, is an increasingly active research topic within the visual analytics community. While much of this work has focused on cognitive or perceptual bias, most closely related to this paper are methods for selection bias transparency.

In particular, this paper builds upon prior work on bias transparency methods for high-dimensional data~\cite{borland_selection_2020}. We leverage similar distance measures and expand upon those visualization designs to identify the amount of selection bias between data subsets. Our work, however,
goes further to offer 
a user-driven bias mitigation workflow that dynamically reweights samples to minimize bias across a user-selected set of dimensions. This workflow is motivated in part by research in survey statistics showing that visualization can help users understand the sample weighting process and its effect on an analysis result~\cite{zhang_increasing_2018}. 

\subsection{Cohort Selection and Event Sequence Visualization}

The DR techniques presented in this paper have been implemented
within the \textit{Cadence} visual analytics system for cohort selection~\cite{gotz_visual_2020}. Cohort selection is a common task in domains such as medical research~\cite{harris_i2b2t2_2016,murphy_serving_2010} and a frequent application for visual analytics technologies~\cite{malik_cohort_2015, zhang_iterative_2014}. \textit{Cadence} leverages event sequence visualization techniques that have been studied by our own research group~\cite{wongsuphasawat_exploring_2012,gotz_decisionflow:_2014,gotz_adaptive_2016,gotz_visual_2020,guo_eventthread_2018} and the broader visual analytics community~\cite{law_maqui_2018,du_coping_2017,wongsuphasawat_lifeflow:_2011,cappers_exploring_2018,malik_high-volume_2016}. The event sequence analysis capabilities of \textit{Cadence}, however, are not the focus of this paper---the DR methods presented
could be applied equally well to a broad range of visual analytics applications involving cohort selection from high-dimensionsal data. 

\subsection{Visualization Designs that Support Reweighting}

The visualizations implemented in \textit{Cadence} to support DR build upon prior work in areas such as visualizing large hierarchies and sets.

\subsubsection{Hierarchical Visualization}

High dimensional data is often organized in hierarchies or other structures that group related variables. For instance, the medical data analyzed in \textit{Cadence} include large type hierarchies such as ICD-10-CM and SNOMED-CT, which contain hundreds of thousands of unique codes representing different diagnoses, procedures, and medications~\cite{nchs_and_cms_icd_2018,spackman_snomed_1997}. Effective bias transparency involves communicating when and where selection bias occurs in these hierarchies of dimensions, and effective bias mitigation involves correcting for bias and understanding the impact of bias correction on analysis results.

Hierarchical visualization techniques include node-link and implicit/space-filling methods~\cite{schulz_design_2011}. Implicit techniques such as tree maps~\cite{johnson_tree-maps:_1991,shneiderman_tree_1992} and icicle plots~\cite{kruskal_icicle_1983} offer information-dense displays via enclosure or adjacency, suitable for large hierarchies; the split icicle plot variant was developed to communicate the emergence of selection bias~\cite{borland_selection_2020}. Although effective for this purpose, it exhibits limitations common to icicle plots such as difficulties in labeling important regions of the hierarchy and incorporating multivariate data. Inspired by previous work 
that integrates hierarchical and table visualizations to display  multivariate attributes (e.g., ~\cite{nobre_lineage_2019}), Section \ref{sec:icicle_table} describes our approach for extending the split icicle plot by integrating a table to incorporate additional information useful for DR.

\subsubsection{Set-Based Visualization}

The DR process 
computes weights for cohort subgroups
formed by the intersections of sets defined by user-specified reweight dimensions. The most common set visualization methods are Venn and Euler diagrams (e.g.,~\cite{wilkinson_exact_2012}), which represent sets as shapes and intersections as overlapping regions. However, such diagrams often grow difficult to interpret as the number of sets increases, leading to the development of various other set visualization techniques~\cite{alsallakh_visualizing_2014}. UpSet, for example, employs a matrix visualization of set intersections, with bar charts to show frequencies for each set and set intersection~\cite{lex_upset_2014}. To prevent overfitting for poorly represented subgroups and to enable the user to adjust the reweighting to mitigate such problems, set and subgroup frequencies are extremely relevant to DR.  We adopt a design similar to UpSet in the reweight set visualization described in Section~\ref{sec:reweight_set_vis}.

\section{Definitions and Design Requirements}

This section enumerates the requirements that motivate our algorithms and user interface designs and briefly describes the medical data used in our examples. Table \ref{tab:definitions} defines key terms used throughout the paper.

\definecolor{lightgray}{gray}{0.9}
\begin{table}[t]
    \centering
        \small
        \rowcolors{1}{}{lightgray}
        \begin{tabular}{|p{.06\textwidth}|p{.38\textwidth}|}
            \hline
            {\bf Term} & {\bf Definition} \\
            \hline
            Selection %
            bias & The result of creating a data sample for analysis such that it is not representative of the larger group intended to be analyzed\\
            Dimension & A measurable property of an individual data entity. Dimensions in our data include attributes (e.g. race or gender) and medical event types (e.g. heart disease or nicotine dependence)\\
            Shift & A change in distribution for a dimension (e.g. increased heart disease prevalence) when comparing two cohorts\\
            Distance & A measure of dimension shift computed by the statistical difference between two distributions \\
            Weighted distance & The distance for a dimension after reweighting is applied to the individuals in a cohort \\
            High-bias dimension & A dimension with a large distribution shift \\
            Constraint & A dimension used to define a cohort (e.g., via a filter operation) \\
            Reweight dimension & A dimension chosen by the user to be reweighted in the DR process \\
            Baseline cohort & The point of comparison for all other cohorts when inspecting for selection bias (typically the initial cohort) \\
            Focus \newline cohort & A cohort selected by the user for in-depth comparison against the baseline \\
            Subgroup & A set of individuals in a cohort with the same values for each of the reweight dimensions, and thus the same weight when reweighting \\
            \hline
        \end{tabular}
        \vspace{0cm}
        \caption{Definitions for key terms used throughout the paper.}
        \vspace{-0.3cm}
\label{tab:definitions}
\end{table}

\subsection{Design Requirements}
\label{sec:design_requirements}

The challenges of selection bias during interactive visual analysis, as well as the DR methods proposed to address them, are general in nature and relevant across domains and applications. However, the work in this paper and the prototype application discussed in Section~\ref{sec:eval} are motivated by feedback gathered by the research team from experts in the medical domain.

The authors have a long history of research exploring new visual analytics techniques and developing tools for cohort selection and analysis (e.g., \cite{zhang_iterative_2014,gotz_icda:_2012}).  This work includes close collaborations with population health researchers at a large cancer hospital \cite{gotz_adaptive_2017} where issues of selection bias and generalizability of research findings to inform clinical care are key priorities. In addition, some of the authors' most recent work includes studies with clinical researchers that evaluated new visual analytics techniques for managing high dimensionality and identifying selection bias in medical cohort studies  \cite{gotz_visual_2020,borland_selection_2020}.

Based on this first-hand experience developing and evaluating visual analytics tools in collaboration with medical domain experts, four key requirements for bias mitigation were identified, as enumerated below.

\begin{itemize}
    \item[R1.] {\bf Identify relevant dimensions that exhibit high levels of bias.} Users should be able to see which dimensions exhibit high levels of selection bias and understand which of those high-bias dimensions, and groups of dimensions, are most relevant for their analytical question.
\vspace{-0.2cm}
    \item[R2.] {\bf Apply bias correction based on user-selected dimensions.} Users should be able to specify one or more dimensions for bias correction and have the system automatically determine the required sample weighting to perform the correction.
\vspace{-0.2cm}
    \item[R3.] {\bf Understand the effect of bias correction.} This includes two key aspects. First (R3.1), 
    as correcting for a small set of specified dimensions will affect a larger set of dimensions, the user should be made aware of how widespread the effects of reweighting are, where bias was reduced, and where and how much bias remains.
    Second (R3.2), users must understand the effect of bias correction on the visualizations driving their primary analysis.
\vspace{-0.2cm}
    \item[R4.] {\bf Prevent overfitting for poorly represented subgroups.} In some cases the weighted samples used for bias correction can excessively amplify poorly sampled subgroups, similar to the problem of model overfitting. Users must be able to understand when a proposed bias correction poses a risk of overfitting due to limitations in the underlying data, and be able to revise the reweighting configuration appropriately.
\end{itemize}

Requirements R1-R4 motivate the algorithms, visualizations, and user workflow presented in this paper.  Moreover, the use case and domain expert interviews described in Section~\ref{sec:eval} are designed to better understand how well the proposed approach meets these requirements.

\subsection{Data Description}
\label{sec:data_description}

The prototype implementation of the techniques described in this paper are applied to medical data which contain both non-temporal attributes (e.g., gender and race) and time-dependent events (e.g., diagnoses and procedures).  Medical events are represented using widely-used coding systems such as ICD-10-CM \cite{nchs_and_cms_icd_2018} and SNOMED-CT \cite{spackman_snomed_1997} for diagnoses and procedures, respectively. These coding systems include over 300,000 distinct codes organized within hierarchical structures, and the electronic health record data can contain events coded at various levels of details.  For instance, a single patient might be diagnosed with a generic ICD-10-CM code of
\textit{I50:~Heart Failure} at one time and the more specific \textit{I50.32:~Chronic diastolic (congestive) heart failure} at another time.  The hierarchical nature of the coding systems means that a patient with the specific \textit{I50.32} diagnosis would also be considered to have the more generic \textit{I50}. This property highlights the importance of understanding selection bias and how DR corrects for it at different levels of specificity in the event type hierarchies (R1 and R3).

\section{Dynamic Reweighting Workflow}

\begin{figure}[t]
    \centering
    \includegraphics[width=0.40\textwidth]{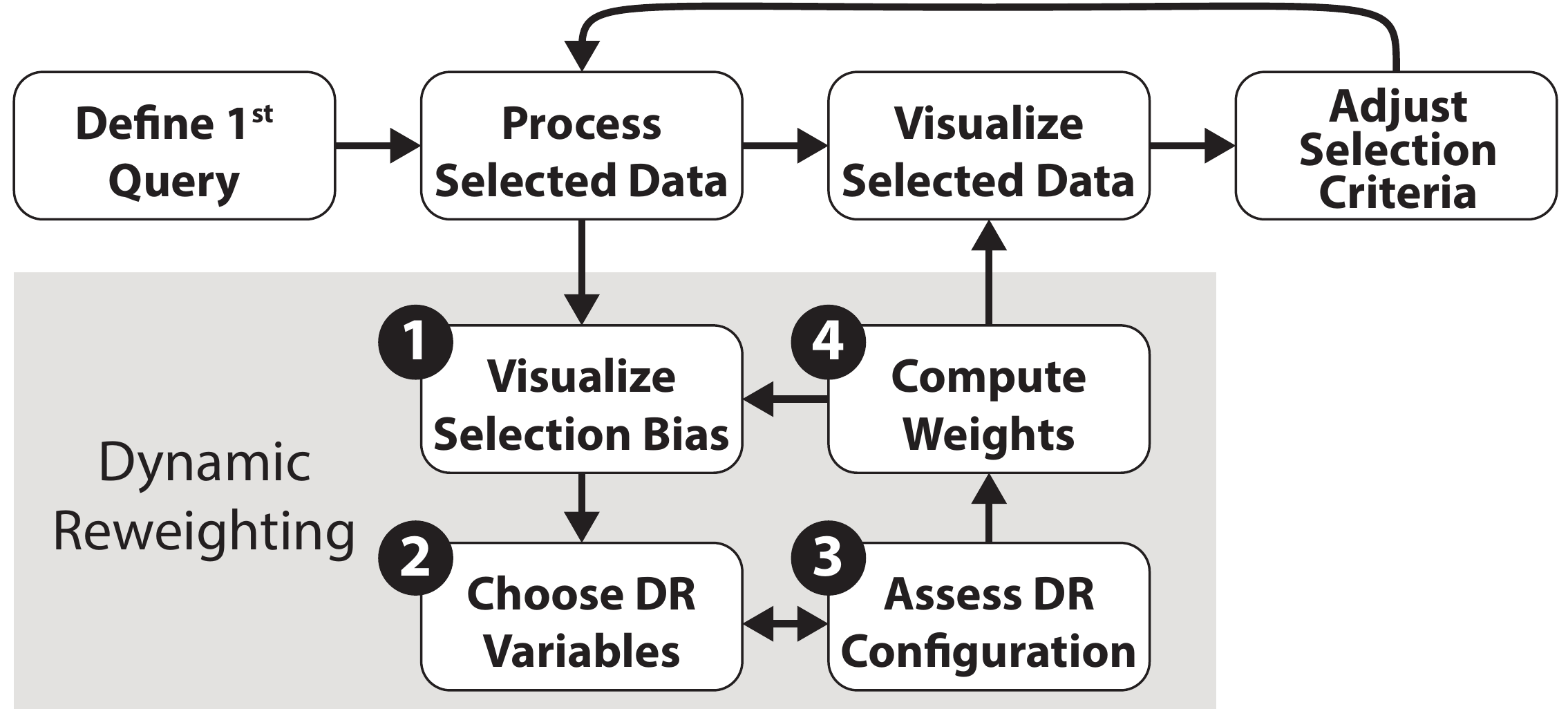}
    \vspace{-0.2cm}
    \caption{Dynamic reweighting (DR) augments the traditional visualization pipeline with a series of additional steps designed to enable user-driven selection bias correction during visualization.}
    \label{fig:workflow}
    \vspace{-0.3cm}
\end{figure}

The dynamic reweighting (DR) workflow augments the canonical visualization process with additional steps that 
enable user-driven bias-corrected visualizations. Traditionally, an input dataset or initial query serves as the initial data for analysis. This data is processed to compute derived values or perform other data transformations, then visualized for users.  Through interaction, users may apply new constraints on selected data to adjust the focus of their analysis. 

The DR workflow adds four additional steps (Figure~\ref{fig:workflow}).  First, DR keeps track of data selections made throughout the visualization process, quantifies the selection bias that emerges due to changing selection criteria, and visualizes this data. This visualization step provides a high-level picture of how much bias was introduced as a side effect of the constraints chosen to define data subsets, as well as detailed information about which dimensions were most affected. Second, users can select specific dimensions to correct for via sample reweighting, so that these dimensions' distributions in the current focus subset more closely approximate the baseline. Note that weights are not applied to the dimensions themselves---they are applied to the individuals in each subset, resulting in distribution shifts across all dimensions. It is typically not possible to fully correct for all distribution differences at the same time, and attempting to do so can excessively amplify individual samples and lead to unreliable results. Users must therefore select a relatively small number of dimensions 
and assess if they would result in a reliable bias correction. This DR configuration assessment is the third stage of the DR workflow.

If the user's assessment suggests that a given DR configuration is problematic, they must revisit the second step to modify the list of selected variables. In contrast, if 
the configuration seems appropriate, 
the user can initiate the fourth and final step in the DR workflow: computing sample weights.  The weights are computed automatically and applied to the originally visualized data
to produce bias-corrected visualizations. The weights also feed back to the selection bias visualizations to 
render an updated view of any remaining bias. 

\vspace{-0.05cm}
\section{Visual Interface Design}
\label{sec:visual_interface}
\vspace{-0.05cm}

\begin{figure}[t]
    \centering
    \includegraphics[width=0.40\textwidth]{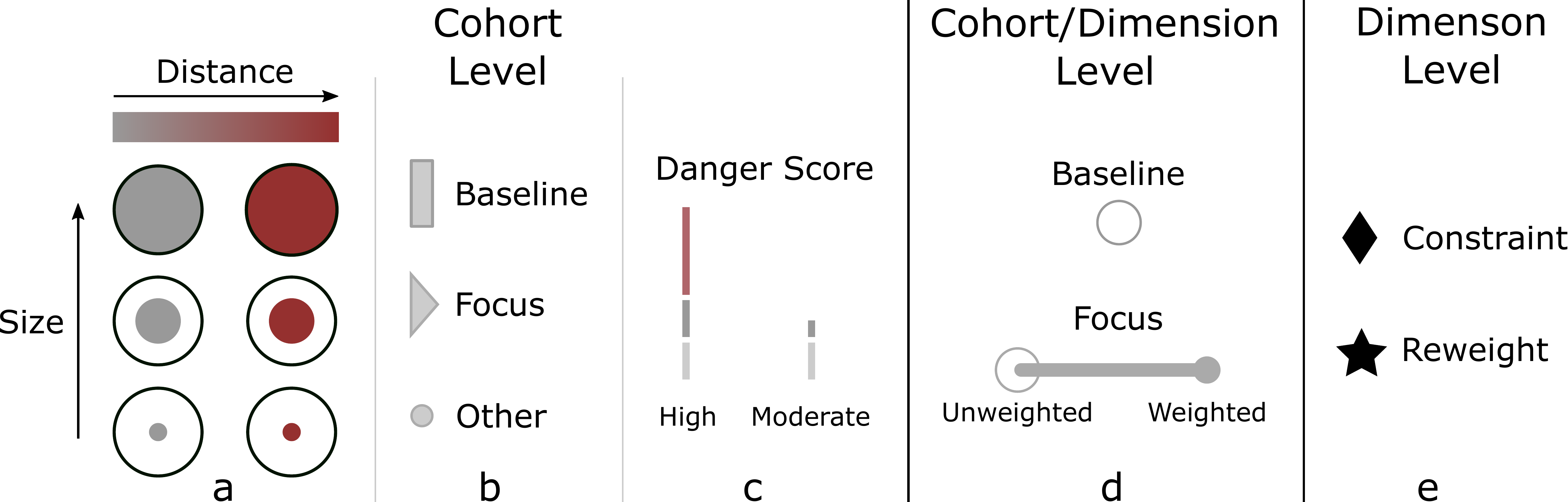}
    \vspace{-0.2cm}
    \caption{Visual interface iconography for cohorts and dimensions.} 
    \label{fig:iconography}
    \vspace{-0.15cm}
\end{figure}

To support the requirements from Section \ref{sec:design_requirements}, we incorporated a number of cohort- and dimension-level visualizations. Much of the iconography is repeated across visualizations to aid in user comprehension (Figure~\ref{fig:iconography}). 
As with~\cite{borland_selection_2020}, the Hellinger distance~\cite{simpson_minimum_1987} is to measure the distribution shift for each dimension.

\vspace{-0.05cm}
\subsection{Cohort Tree}
\vspace{-0.05cm}

\begin{figure}[t]
    \centering
    \includegraphics[width=0.45\textwidth]{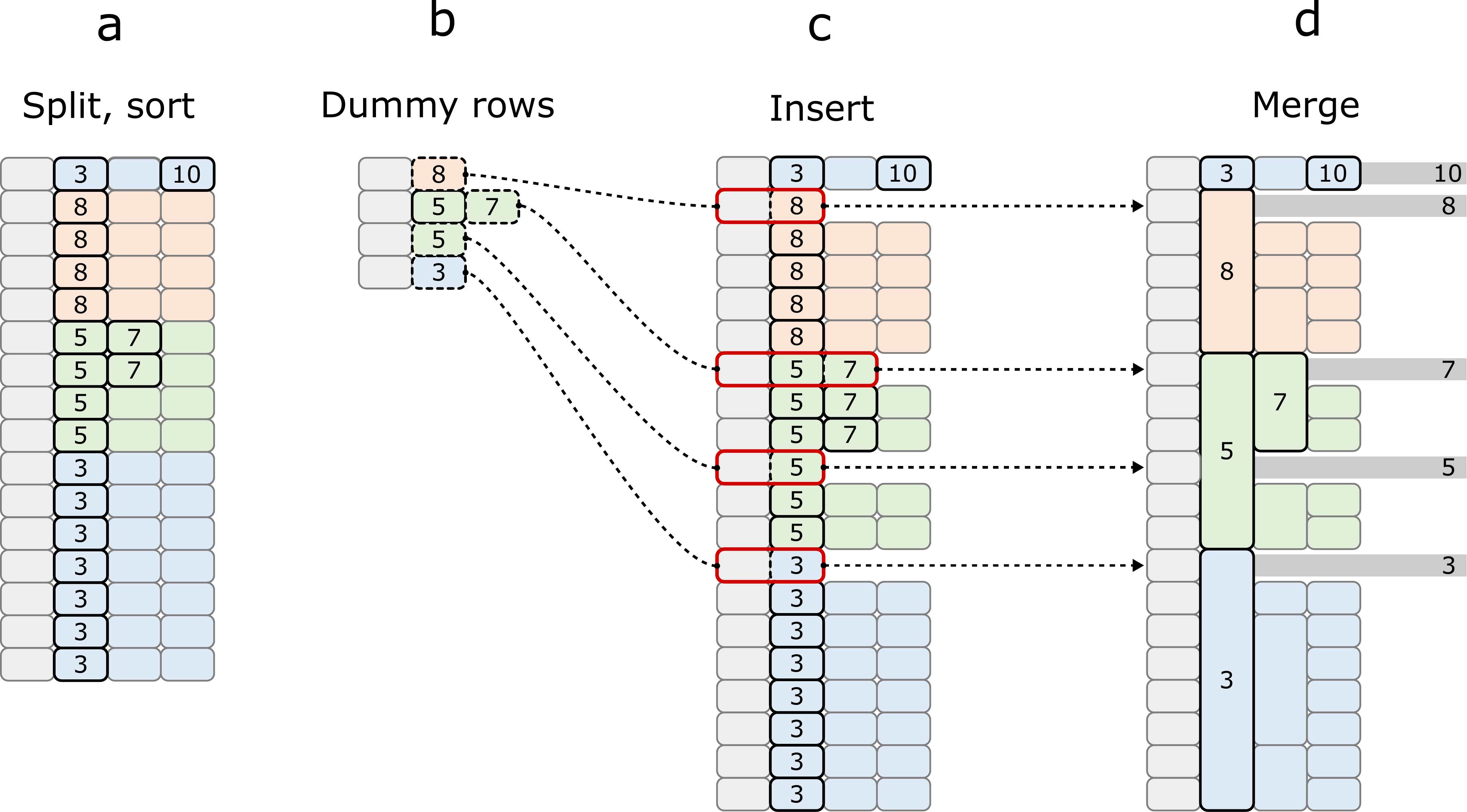}
    \vspace{-0.2cm}
    \caption{(a) Icicle table construction begins with a split and sorted icicle plot. (b) Dummy rows are created for each non-leaf salient node, and then (c) inserted into the icicle plot. (d) Adjacent nodes are then merged, leaving room for labels and the insertion of rows into a table visualization.} 
    \label{fig:icicle_table_algorithm}
    \vspace{-0.2cm}
\end{figure}

As the event sequence analysis capabilities of \textit{Cadence} are used to filter existing cohorts to create new cohorts, representations of each cohort and their provenance are shown in the cohort provenance tree (Figure~\ref{fig:teaser}-a). Each cohort is represented by a glyph encoding cohort size and aggregate distance across all data dimensions, along with icons indicating the baseline and focus cohorts (Figure~\ref{fig:iconography}-a and b). This design is similar to that described in~\cite{borland_selection_2020}; however, two crucial extensions have been made to support the DR workflow: (1)~a \textit{danger score} indicator has been added, and (2)~a new aggregate distance measure is used to indicate potential selection bias.

The danger score alerts the user to potential problems with the current reweighting configuration (R4). A score is computed for each cohort (Section \ref{sec:danger_score}) to identify those with poorly sampled subgroups. An indicator is shown next to any cohort with a score approaching a system-defined threshold, indicating  that adjustments to the current reweighting configuration may be warranted (Figure~\ref{fig:iconography}-c).

An improved aggregate distance measure is included to better summarize the effect of reweighting upon a cohort (R3.1). In a typical cohort exhibiting selection bias, the vast majority of dimensions undergo small to moderate shifts in distribution, whereas a smaller number of dimensions that are highly correlated with the dimensions used for selection will undergo larger shifts. It is typically these high-bias dimensions that the user wishes to correct. In~\cite{borland_selection_2020} the mean Hellinger distance across all dimensions was used as an aggregate selection bias measure for each cohort. However, a successful reweighting that reduces bias in dimensions with large shifts will likely not have much effect on the mean, as these dimensions constitute a small fraction of all dimensions. To indicate whether a reweighting was successful, we therefore use a metric that more heavily reflects dimensions with large shifts, the generalized mean (a.k.a. power mean), which is defined as
\vspace{-0.18cm}
\begin{equation}
    M_p(x_1,\ldots,x_n) = \left(\frac{1}{n} \sum_{i=1}^{n} x^{p}_{i}\right)^{\frac{1}{p}}
\vspace{-0.18cm}
\end{equation}
where $p$ is a non-zero real number and $x_1,...,x_n$ are real numbers $\geq 0$. Increasing $p$ weights larger values more than smaller values (with $M_1$ equivalent to the arithmetic mean, and $M_\infty$ producing $\max(x_1,\ldots,x_n)$). We have chosen $M_8$ as a value that works well for our current data.

\subsection{Icicle Table}
\label{sec:icicle_table}

The split icicle plot was developed to visualize shifts in distribution for dimensions in large hierarchies, indicating potential selection bias~\cite{borland_selection_2020}. Although effective for communicating this information, it exhibits some limitations when used for DR. We therefore developed the icicle table to address these limitations and add functionality useful for DR (R1-R3).

The split icicle plot modifies the strict hierarchical icicle plot layout by splitting certain nodes, enabling more effective sorting of the plot by the maximum distance along each path from a leaf node to the root. Thus areas with large shifts can be sorted toward the top of the plot to help the user prioritize dimensions to investigate for selection bias (R1). In addition, aggregation is used to reduce the number of rendered nodes and emphasize ``salient'' nodes that indicate areas where selection bias is rapidly increasing or decreasing.

One critical limitation of the split icicle plot, and icicle plots in general, is the difficulty of
labeling nodes due to the plot's dense layout. In~\cite{borland_selection_2020} the user could obtain information on any node via mouseover. Still, users indicated that the ability to understand at a glance which portions of the hierarchy 
had large distribution shifts was crucial in interpreting the data. Below we describe modifications to the split icicle plot layout that enable (1) the placement of labels for salient nodes at any level in the hierarchy and (2) the creation of rows for these nodes in a table visualization to display additional multivariate information. Throughout this paper, we use the term ``icicle table'' to refer to the combined split icicle plot/table visualization, and ``icicle plot'' and ``table'' to refer to their respective components.

\subsubsection{Layout Algorithm}

Figure~\ref{fig:icicle_table_algorithm} shows the construction of an icicle table layout. In (a), a split icicle plot after the split and sort steps is shown, displaying the distance values for all salient nodes. In (b), a ``dummy'' row is created for each non-leaf salient node, comprising that node and its ancestors (dummy nodes are not necessary for salient leaf nodes, which already have space to their right for labels and table rows). In (c), the sorted dummy nodes are inserted from top to bottom. Finally, adjacent nodes are merged in (d). The insertion of dummy rows for non-leaf salient nodes provides room for labels and table rows in an integrated table visualization.
 
\subsubsection{Icicle Plot Visualization}

\begin{figure}[t]
    \centering
    \includegraphics[width=0.47\textwidth]{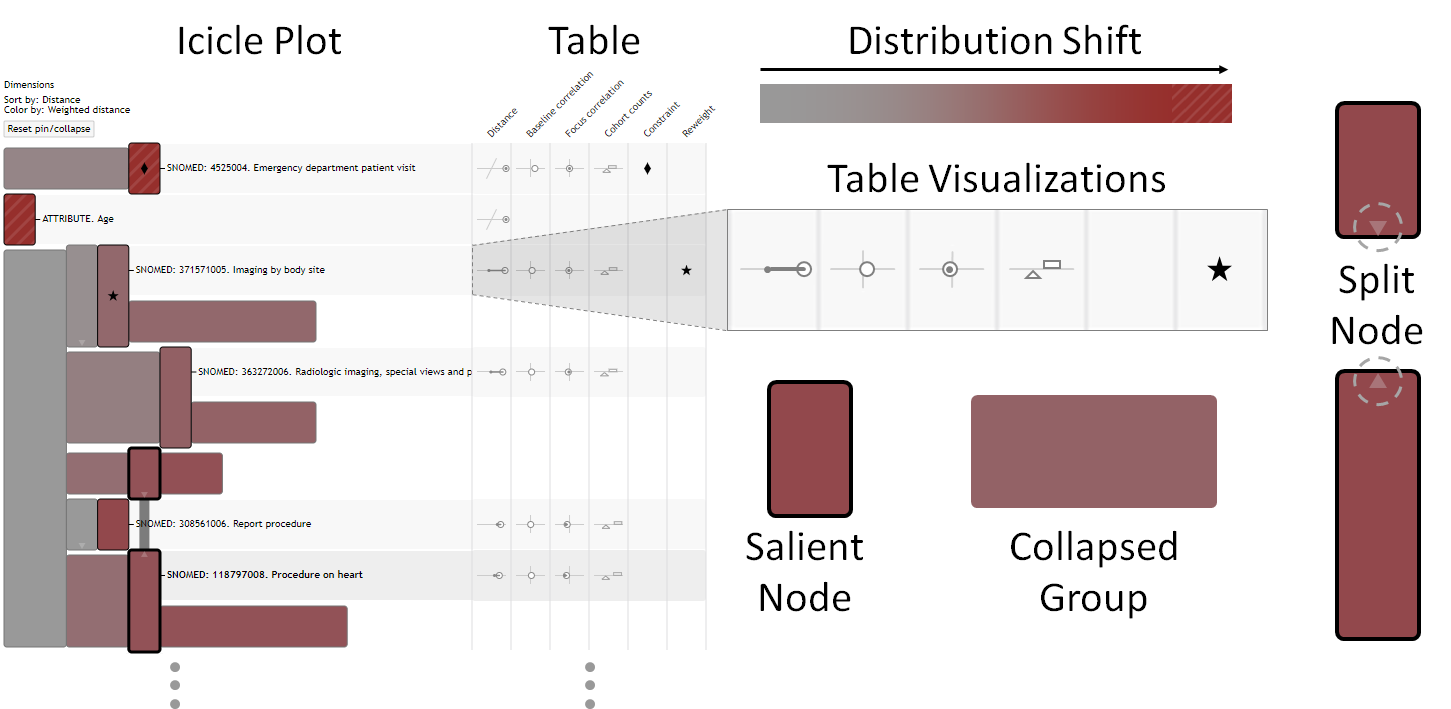}
    \vspace{-0.2cm}
    \caption{Example icicle table. The split icicle plot layout (left) is modified such that salient nodes have room for labels and rows in a table visualization (right). Multiple attributes for each dimension, such as distance and correlations with outcome, are displayed in the table.} 
    \label{fig:icicle_table_example}
    \vspace{-0.1cm}
\end{figure}

Figure~\ref{fig:icicle_table_example} shows an example icicle table. By default, each node is colored by its distribution shift and each collapsed group by the maximum shift in the group. Collapsed groups can be expanded on demand to show a standard icicle plot of the contained dimensions. In~\cite{borland_selection_2020}, separate plots were used for each event type hierarchy (ICD-10-CM and SNOMED-CT), and attributes (e.g., race and gender) were visualized separately. However, to identify dimensions useful for DR, it is better to display all dimensions in the same plot to emphasize which have shifted the most. 
 A single layout is therefore calculated for all dimensions, and attributes are categorized as salient to ensure their visibility.

Constraint dimensions for the baseline and focus cohorts are indicated with a $\blacklozenge$, and reweight dimensions with a $\bigstar$ (Figure~\ref{fig:iconography}-e). The color map is scaled to the maximum distribution shift for all dimensions that are neither constraints nor descendants of a constraint, as constraints were deliberately chosen by the user to undergo a distribution shift. Hatched lines indicate nodes with a value greater than the color map maximum.

\begin{figure}[t]
    \centering
    \includegraphics[width=0.47\textwidth]{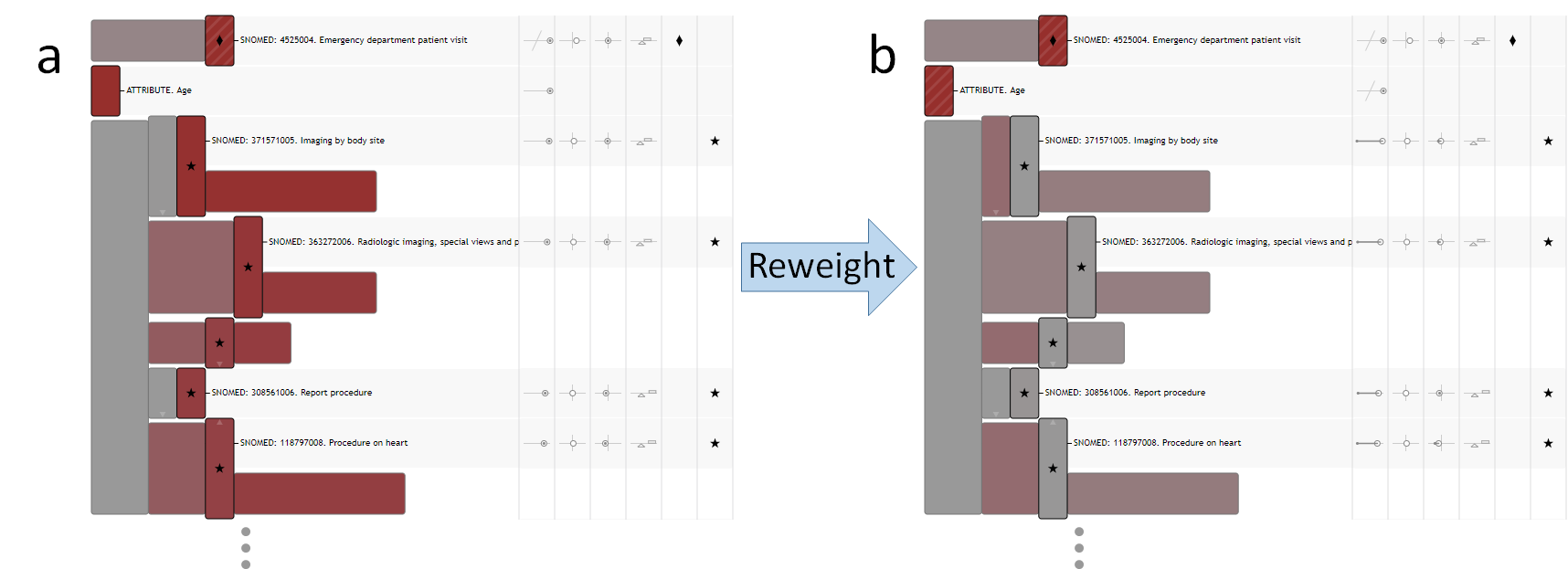}
    \vspace{-0.2cm}
    \caption{Icicle table sorted by unweighted distance and colored by weighted distance. (a) No reweighting applied. (b) Reweighting applied. The changes from red in (a) to grey in (b) indicate that the reweighting process has successfully corrected for shift in several dimensions.} 
    \label{fig:icicle_table_reweight}
    \vspace{-0.25cm}
\end{figure}

The gradient-based saliency metric from~\cite{borland_selection_2020} captures areas of the dimension hierarchy where the shift in distribution increases or decreases rapidly, but it suffers from two limitations when applied to DR: (1) it does not capture situations where drifts accumulates gradually, and (2) it was designed for a distance metric with values $\geq 0$. We therefore use a modified saliency criterion, given shift gradient $\Delta s_i$ defined as the change in distribution shift $s_i$ between dimension $i$ and its parent. First, all dimensions with $\left| \Delta s_i \right| \geq t_s$, where $t_s$ is a user-defined saliency threshold, are marked as salient. Then, for any path from a leaf node to the root containing a dimension with $|s_i| \geq t_s$ that does not contain a salient dimension, the dimension with the greatest $\left| \Delta s_i \right|$ is marked salient. Thus, all areas of the hierarchy with large shifts receive a label and a row in the table, and the layout can easily incorporate metrics, such as correlation, that can have negative values.

Setting $t_s$ enables high-level control of the aggregation level of the icicle plot, however the user can also fine-tune the layout using dimension \textit{pin} and \textit{collapse} controls. Pinning a dimension classifies it as salient regardless of $t_s$, enabling the user to mark dimensions known a priori to be of interest or to mark candidate reweight dimensions as they explore the data. Collapsing a dimension classifies the dimension and all of its descendants as non-salient, regardless of $t_s$. This feature can deemphasize dimensions that the user knows are not of interest for reweighting. For example, selecting $Gender = Female$ may cause dimensions related to pregnancy to shift dramatically, dominating the visualization. The user can collapse such dimensions to focus on others. 

To further support the DR workflow, we have added a \textit{replace reweight} mode (R2 and R4). When fine-tuning the reweighting configuration, the user may wish to adjust a reweight dimension by moving up in the hierarchy to a more general type or down to a more specific type. To do so, the user can enter replace reweight mode, causing the icicle plot to show only the reweight dimension, its ancestors, and its descendants. All ancestors, the reweight dimension, and two levels of children are marked as salient, providing a detailed view of the local neighborhood around the reweight dimension. The user can then select a dimension to take its place in the list of reweight dimensions.

\vspace{-0.05cm}
\subsubsection{Table Visualization}
\vspace{-0.05cm}

Integrating a table with the split icicle plot enables the inclusion of multi-attribute information
to help the user select appropriate dimensions for reweighting (R1). To support the DR process we include \textit{Distance}, \textit{Baseline correlation}, \textit{Focus correlation}, \textit{Cohort counts}, \textit{Constraint}, and \textit{Reweight} columns that display the table visualizations shown in Figure~\ref{fig:icicle_table_example}.
Correlation with outcome visualizations help prioritize dimensions that are closely related to the outcome of interest. 
These visualizations use the iconography shown in Figure~\ref{fig:iconography}-b, d, and e.

The table can also
be used to adjust the 
appearance of the icicle plot. Column context menus enable the user to sort the layout and color by (a) unweighted distance, (b) weighted distance, (c) baseline correlation, (d) unweighted focus correlation, or (e) weighted focus correlation. By default the plot is sorted by (a) and colored by (b), enabling the user to easily see the effects of reweighting via change in color (Figure~\ref{fig:icicle_table_reweight}). However, the user can select different combinations of distance and correlation measures to explore 
in more detail.
For (c-e), magnitude is used for sorting, and a double-ended blue-grey-red map for coloring.

\vspace{-0.05cm}
\subsection{Distance vs. Correlation Plots}
\label{sec:distance_correlation_plots}
\vspace{-0.05cm}

\begin{figure}[t]
    \centering
    \includegraphics[width=0.47\textwidth]{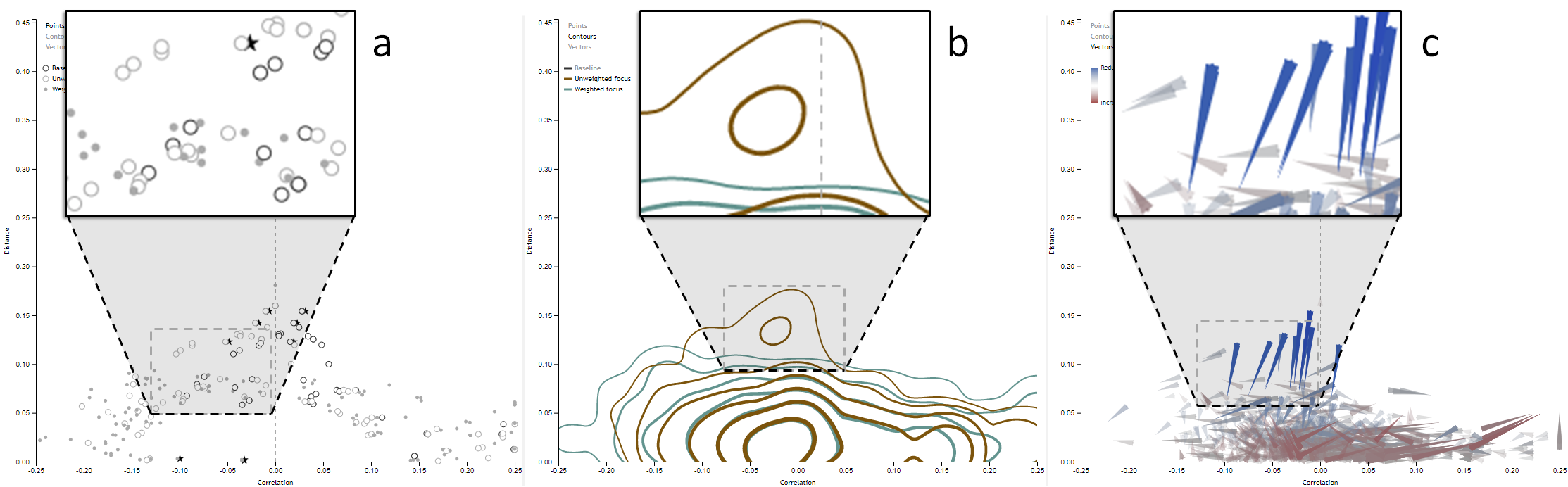}
    \vspace{-0.25cm}
    \caption{Distance vs. correlation plots: (a) scatter plot, (b) contour plot (c) vector plot. These plots enable the user to see per-dimension changes in distance and correlation with outcome due to reweighting, as well as overall shifts in these distributions.} 
    \label{fig:distance_correlation}
    \vspace{-0.25cm}
\end{figure}

Three additional visualizations show the effect of reweighting on per-dimension distances and outcome correlations for the baseline and focus cohorts (R3) and enable the selection of reweighting dimensions (R1): a scatter plot, contour plot, and vector plot (Figure~\ref{fig:distance_correlation}). Each view shows correlation with outcome along the x-axis and focus-to-baseline distance along the y-axis. The user can switch views on demand, with linked selection between the three views and the icicle table.

\vspace{-0.05cm}
\subsubsection{Scatter Plot}
\vspace{-0.05cm}

The scatter plot (Figure~\ref{fig:distance_correlation}-a) enables the user to select individual dimensions based on distance and correlation with outcome, and to see relationships between the baseline, focus, and weighted focus values for each dimension. To prevent overplotting with tens of thousands of dimensions, points are filtered based on local density, which typically renders those with high distance and/or correlation values. Three glyphs, representing values for the baseline, unweighted focus, and weighted focus cohorts, are rendered for each dimension using the icons from Figure~\ref{fig:iconography}-d. Display of each glyph type can be toggled by the user. Connections between glyphs of the same dimension are shown on mouseover or selection. 
    
\vspace{-0.05cm}
\subsubsection{Contour Plot}
\vspace{-0.05cm}

The contour plot shows overall differences in distribution between the baseline, unweighted focus, and weighted focus cohorts. Colored contours are drawn for each cohort, and the user can toggle which cohorts to display. Figure~\ref{fig:distance_correlation}-b shows an example where the overall distance in the focus cohort (brown) has been reduced in the weighted version (green), and changes in correlation have also occurred.

\subsubsection{Vector Plot}

The vector plot (Figure~\ref{fig:distance_correlation}-c) visualizes changes in the focus cohort due to reweighting. The vector base represents its unweighted value and points towards the position of the weighted value.
Length and opacity are scaled by vector magnitude and colored by distance with a blue (reduced distance) to red (increased distance) color map. Vectors are filtered by vector magnitude to reduce overplotting. This visualization 
shows
which dimensions have shifted the most and in what direction.

\subsection{Dimension Distribution Plots}

\begin{figure}[t]
    \centering
    \includegraphics[width=0.47\textwidth]{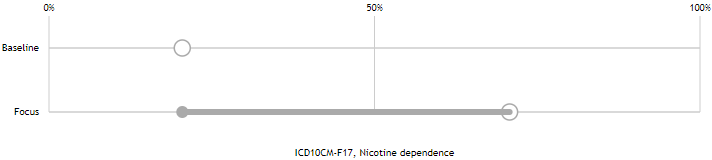}
    \vspace{-0.2cm}
    \caption{Dimension distribution plot. Reweighting the focus cohort reduces nicotine dependence prevalence, matching the baseline.} 
    \label{fig:event_distribution}
    \vspace{-0.2cm}
\end{figure}

Data type-dependent visualizations supporting R3 show the distribution of any dimension selected via the icicle table or distance vs. correlation plots, enabling the user to view information such as which cohort has a higher percentage of heart disease or what the gender breakdown is for each cohort. Figure~\ref{fig:event_distribution} shows the distributions of nicotine dependence in the baseline (top) and focus (bottom) using the designs from Figure~\ref{fig:iconography}-d. In this example, the weighted focus cohort's distribution has shifted to match that of the baseline cohort. This design is also incorporated into a histogram visualization for numeric dimensions (e.g., age) and a dumbbell plot for categorical dimensions (e.g., gender or race).

\subsection{Balance Interface}
\label{sec:balance_intervace}

The balance panel (Figure~\ref{fig:teaser}-b) includes visualizations and controls for the DR process, supporting R2 and R4. A reweight list shows all dimensions selected for reweighting. The user can remove dimensions from the list, initiate the reweighting process, and use a slider to control the amount of reweighting. A detailed view of the per-cohort subgroups used for reweighting is shown in the reweight set visualization.

\subsubsection{Reweight Set Visualization}
\label{sec:reweight_set_vis}

The reweight set visualization (Figure~\ref{fig:teaser}-c) is based on Upset~\cite{lex_upset_2014}. It shows how the reweight dimensions combine to form subgroups of each cohort to be reweighted (Section~\ref{sec:algorithms}), and enables identification and correction of potential reweighting issues 
that may result in unreasonably high weights for small subgroups of individuals (R4).

Each column represents a reweight dimension, and each row a subgroup formed by combinations of these dimensions. As the reweight dimensions in our system are exclusively binary event types, for $n$ dimensions there will be $2^n$ subgroups (including any empty subgroups). Linear plots above each column indicate the frequency of each event type per cohort. Cohorts are indicated as shown in Figure~\ref{fig:iconography}-b. To the right of each row, a similar visualization indicates the corresponding subgroup size for each cohort. Rows are sorted in ascending order by average subgroup size, such that potentially problematic subgroups are located toward the top of the table. An integrated visualization also shows the danger score for each cohort, colored red if above the danger score threshold. 
If the user determines that a particular dimension should be adjusted to correct a reweighting problem, they can enter the icicle table's replace reweight mode (Section~\ref{sec:icicle_table}) by selecting it from a context menu available for each dimension.

\section{Reweighting Methods}
\label{sec:algorithms}

\definecolor{lightgray}{gray}{0.9}
\begin{table}[t]
    \centering
        \small
        \rowcolors{1}{}{lightgray}
        \begin{tabular}{|p{.06\textwidth}|p{.38\textwidth}|}
            \hline
            {\bf Symbol} & {\bf Definition} \\
            \hline
            $\mathbf{B}$,$\mathbf{F}$ & Baseline \& focus cohorts respectively\\
            $B,F$ & Total number of entities in the baseline \& focus cohorts respectively\\
            $B_i,F_i$ & Total number of entities in subgroup $i$ for the baseline \& focus cohorts respectively\\
            $w_i,w_{i,interp}$ & Weight and interpolated weight for subgroup $i$ \\
            $C$ & Interpolation coefficient \\
            $E_{Bi},E_{Fi}$ & Expected counts for subgroup $i$ in baseline \& focus cohorts respectively \\
            $D_k$ & Danger score for $k$ subgroups \\
            $D$ & Standardized danger score \\
            \hline
        \end{tabular}
        \vspace{-0cm}
        \caption{A summary of the notation used in Section \ref{sec:algorithms}.}
        \vspace{-0.3cm}
\label{tab:notation}
\end{table}

Using the visualizations described in Section \ref{sec:visual_interface}, the user can choose the dimensions to correct for via reweighting. Currently only event types, which are binary, can be selected, although the reweighting methods described below support other data types such as categorical or binned numeric attributes.
The reweighting process is described using a baseline and single focus cohort, however in practice each non-baseline cohort in the system is reweighted with respect to the baseline using the same set of reweight dimensions to produce comparable visualizations. Subgroups of each cohort are created based on combinations of the reweight dimensions, and the reweighting process computes weights for each subgroup. These weights are applied to each entity in the subgroup and 
used to compute weighted statistics for each cohort.

\subsection{Weight Calculation and Interpolation}
\label{sec:weight_calculation}

Consider a baseline cohort $\mathbf{B}$ with $B$ entities and a focus cohort $\mathbf{F}$ with $F$ entities. Given $n$ binary dimensions, $\mathbf{B}$ can be divided into $k$ subgroups where $0<k\leq 2^n$. A strict equality of $k=2^n$ does not hold as empty subgroups may exist in both $\mathbf{B}$ and $\mathbf{F}$.\footnote{In the more general case of $n$ categorical dimensions with $c_j$ categories for dimension $j$, $k$ would be bound by $\prod_{j=0}^{n-1}c_j$.} Each subgroup $i$, $0 \leq i <k$, of $\mathbf{B}$ and $\mathbf{F}$ will have size $B_i$ and $F_i$ respectively. If $F_i \neq 0$ for all $0 \leq i <k$, we define weights for subgroup $i$ in $\mathbf{F}$ as
\vspace{-0.15cm}
\begin{equation} \label{weight_eq1}
    w_i=\frac{B_iF}{F_iB}
\vspace{-0.15cm}
\end{equation}
The intuition behind Equation \ref{weight_eq1} is that if every entity in $\mathbf{F}$ receives the weight computed for its respective subgroup, the proportion of each subgroup relative to the total cohort size will be the same in both the  $\mathbf{B}$ and  $\mathbf{F}$ subgroups. If $\exists F_i=0$ for some $0 \leq i <k$, we instead define weights for subgroup $i$ as
\vspace{-0.15cm}
\begin{equation} \label{weight_eq2}
    w_i=\frac{B_iF}{F_i\sum_{j=0}^{k-1}B_jI(F_j \neq 0)}
\vspace{-0.15cm}
\end{equation}
where $I(\cdot)$ is the indicator function. The intuition behind Equation \ref{weight_eq2} is the same as for Equation \ref{weight_eq1}, except that after matching the proportions we adjust the weights so that the size of the weighted cohort is the same as that of the corresponding unweighted one. The weight can also be interpolated to achieve a balance between weighted and unweighted values.
For any constant $C\in [0,1]$, where $0$ applies no reweighting and $1$ full reweighting, the interpolated weights are $w_{i,interp}=1+C(w_i-1)$.

\subsection{Danger Score Calculation}
\label{sec:danger_score}
\definecolor{lightgray}{gray}{0.9}
\begin{table}[t]
    \centering
        \small
        \rowcolors{1}{}{lightgray}
        \setlength\tabcolsep{0.8em}
        \begin{tabular}{|c|cccc|cccc|c|c|}
            \hline
            \multicolumn{1}{|l|}{} &
            \multicolumn{4}{c|}{\bf B Subgroup Sizes} & 
            \multicolumn{4}{c|}{\bf F Subgroup Sizes} & 
            \multicolumn{1}{c|}{\bf $D_k$} &  
            \multicolumn{1}{c|}{\bf $D$}\\
            \hline
            1 & 100 & 200 & 300 & 400 & 0 & 200 & 300 & 400 & 95 & 85.59\\
            2 & 100 & 200 & 300 & 400 & 0 & 2 & 3 & 4 & 0.999 & 0.063\\
            3 & 1 & 200 & 300 & 400 & 0 & 200 & 300 & 400 & 0.009 & $1.1\times 10^{-7}$\\
            \hline
        \end{tabular}
        \vspace{-0.05cm}
        \caption{Subgroup sizes and danger scores used in examples. }
        \vspace{-0.3cm}
\label{tab:danger_examples}
\end{table}

Using this weighting framework we can calculate weights for any $\mathbf{B}$ and $\mathbf{F}$ regardless of their subgroup distributions. An important question to consider, however, is whether or not a reweighting is appropriate. If $\mathbf{B}$ and $\mathbf{F}$ have drastically different subgroup distributions, a reweighting may not be appropriate because the populations might be fundamentally different with respect to the reweight dimensions. We therefore developed a reweight configuration \textit{danger score} based on the chi-square statistic to assess the appropriateness of any reweighting.

Consider a table of subgroup sizes with two rows---one for $\mathbf{B}$ and one for $\mathbf{F}$---and $k$ columns---one per subgroup. The row sums would be $B$ and $F$, and the column sums would be $S_i=B_i+F_i$, $0 \leq i <k$. The expected counts for subgroup $i$ in $\mathbf{B}$ and $\mathbf{F}$ are defined as
\vspace{-0.1cm}
\begin{equation} \label{danger_eq1}
    E_{Bi}=\frac{S_iB}{B+F} \quad\mathrm{and}\quad E_{Fi}=\frac{S_iF}{B+F}
\vspace{-0.1cm}
\end{equation}
Using $E_{Bi}$ and $E_{Fi}$, we construct the danger score for $k$ subgroups $D_k$ from the chi-square statistic for independence:
\begin{equation}
    D_k=\sum_{i=0}^{k-1}\frac{(B_i-E_{Bi})^2}{E_{Bi}}+\sum_{i=0}^{k-1}\frac{(F_i-E_{Fi})^2}{E_{Fi}}
\end{equation}
As the subgroup sizes of $\mathbf{B}$ and $\mathbf{F}$ start to differ from their expected values, $D_k$ will increase~\cite{agresti_categorical_2002}. Therefore a high value of $D_k$ indicates that reweighting could be inappropriate, as the populations of $\mathbf{B}$ and $\mathbf{F}$ could be fundamentally different for reasons other than biased sampling.

To compare danger scores for different values of $k$ we need to standardize the scores with respect to their asymptotic tail probabilities. As sample sizes increase, $D_k$ converges to a $\chi^2_{k-1}$ distribution~\cite{agresti_categorical_2002}. Therefore we can find the standardized danger score $D$ of $D_k$ as
\begin{equation}
    D=F^{-1}_1(F_{k-1}(D_k))
\end{equation}
where $F_{k}(x)=P(X^2_{k}<x)$ and $X^2_k\sim\chi^2_k$. The critical chi-square value for one degree of freedom and a $p$-value of $0.05$ is $3.84$, so if $D<3.84$ there is no evidence for statistical dependence between the reweight dimensions and the criteria for selecting $\mathbf{F}$. In this case all deviations from the expected subgroup sizes are due to random variability. Note that the difference between values of $D$ can only be interpreted from the probabilistic statement, which is non-linear. A value of $D$ twice as large does not imply that the reweighting is twice as inappropriate.

Very large values of $D_k$ can occur when $\mathbf{B}$ and $\mathbf{F}$ have substantial heterogeneity with respect to the reweight dimensions. This poses a problem when calculating 
$D$, as $F_k(x)$ is computed numerically. A large $D_k$ results in $F_k(D_k)$ being very close to $1$, and computing $F^{-1}_1(F_k(D_k))$ would produce a numerical result of $\infty$. Although such a large value would already be a warning to the user that this particular reweighting is inappropriate, it is desirable to retain the comparability of all possible danger scores. After conducting numerical comparisons of values of $D_k$ and $D$, we found that at values of $D>50$, the relationship between $D_k$ and $D$ is close to linear for all $k\geq 2$. Furthermore, since computational methods start to fail at $D>70$, we use this linearity to approximate the standardization process at larger values. We find a computational limit $L_k$ for $D_k$ as
\begin{equation}
    L_k=F^{-1}_{k-1}(F_{1}(L))
\vspace{0.01cm}
\end{equation}
where $L=50$. If $D_k>L_k$ we use the following approximation for $D$:
\begin{equation}
    D=\frac{L-F^{-1}_1(F_{k-1}(L_k-\epsilon))}{\epsilon}(D_k-L_k)+L
\end{equation}
for any reasonable $\epsilon>0$. In our implementation $\epsilon=5$ was used. Intuitively, as the distributions of subgroups start to differ between $\mathbf{B}$ and $\mathbf{F}$, $D_k$ will increase. As an example, consider subgroup sizes in row one of Table~\ref{tab:danger_examples}. Since there exists a subgroup with a stark contrast between $\mathbf{B}$ and $\mathbf{F}$ (the first subgroup in each), this difference results in a large $D_k$. Now consider the subgroup sizes in row two. This configuration results in a very small $D_k$. Intuitively, although the distribution of subgroup sizes in $\mathbf{F}$ in the second row is proportional to that in the first, the smaller $F$ means that the zero size for the first subgroup could have arisen from expected variability or sampling bias and is not extreme enough for the user to be concerned with when reweighting. Finally, consider the last row of Table~\ref{tab:danger_examples}. In this example $D_k$ will also be very small, as the zero size for the first subgroup in $\mathbf{F}$ is within the expected variability of the size distribution for the first subgroup in $\mathbf{B}$. Empirically, we determined that a value of $D>50$ was reasonable for warning the user.  The appropriate warning threshold, however, may vary based on data characteristics. To make this value more intuitive and easier to interpret, $D$ is divided by a warning threshold of $50$, resulting in a normalized danger score threshold of one. 
 
\subsection{Weighted Statistics}
\label{sec:weighted_stats}

Reweighting assigns each entity in a cohort a weight based on its subgroup. Aggregate cohort statistics, such as outcome rates and event prevalences, can then be recalculated to account for these weights. 
For binary correlation with an outcome we use the weighted Pearson correlation coefficient~\cite{bailey_weighted_2018}. Consider $n$ entities with binary events $V={v_1,v_2,...,v_n}$, $n$ binary outcomes $O={o_1,o_2,...,o_n}$, and $n$ weights $\Omega={\omega_1,\omega_2,...,\omega_n}$. The weighted correlation $\rho_w$ is
\begin{equation}
    \resizebox{.44\textwidth}{!}{%
        $\rho_w=\frac{\sum_{i=1}^n\omega_i\left(v_i-\frac{\sum_{j=1}^n \omega_j v_j}{\sum_{j=1}^n\omega_j}\right)\left(o_i-\frac{\sum_{j=1}^n \omega_j o_j}{\sum_{j=1}^n\omega_j}\right)}{\sqrt{\sum_{i=1}^n\omega_i\left(v_i-\frac{\sum_{j=1}^n \omega_j v_j}{\sum_{j=1}^n\omega_j}\right)^2}\sqrt{\sum_{i=1}^n\omega_i\left(o_i-\frac{\sum_{j=1}^n \omega_j o_j}{\sum_{j=1}^n\omega_j}\right)^2}}$%
    }
\end{equation}

The interpolated correlation can be calculated directly using the interpolated weights. However, this approach is computationally expensive when applied to all dimensions in our data.
We therefore compute a linear interpolation between the unweighted and weighted correlations, $\rho_{interp}=\rho(1-C)+C\rho_W$, for any $C\in [0,1]$.
For our data and use case the approximation was plausible, however exact values could be computed if desired.
\section{Example Use Case and Domain Expert Interviews}
\label{sec:eval}

\begin{figure*}[t]
    \centering
    \includegraphics[width=0.874\textwidth]{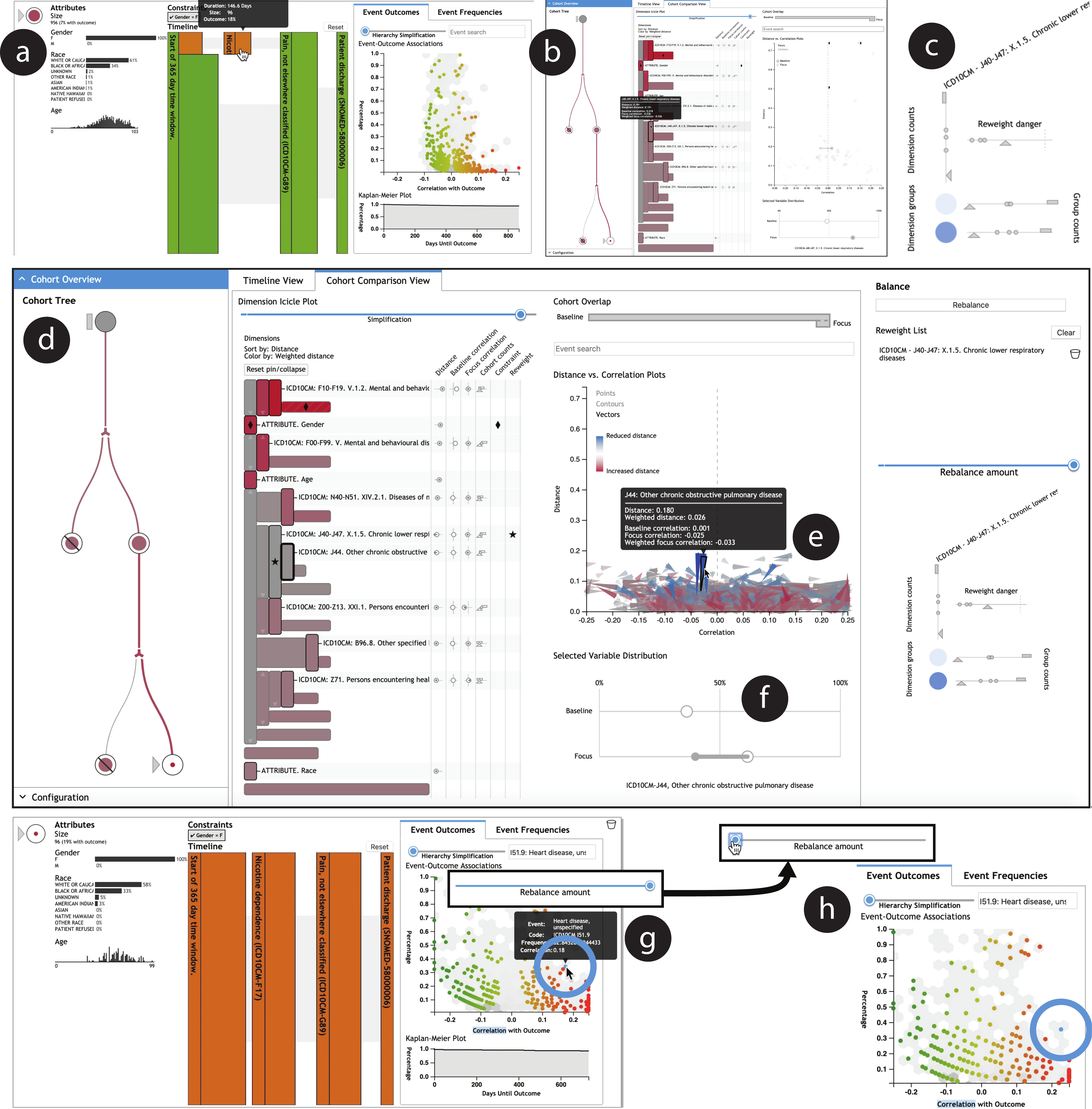}
    \vspace{-0.2cm}
    \caption{This use case for DR demonstrates how reweighting to correct for selection bias can have large effects on resulting visualization. In this case, the correlation between a Heart Failure diagnosis and Opiate Disorders for the target subgroup was reduced by reweighting from over 0.2 to about 0.18, and the diagnosis no longer stood out as an unusually high risk factor.}
    \label{fig:scenario}
    \vspace{-0.2cm}
\end{figure*}

This section describes a real-world use case from the medical domain that demonstrates both the user workflow supported by DR and its impact on the visual analysis scenario. It also presents a thematic analysis of comments gathered from semi-structured interviews with health researchers.  Both the use case and domain expert interviews report DR as implemented within the \textit{Cadence} visualization system \cite{gotz_visual_2020,borland_selection_2020}. 
While participants commented on a broad range of analytical features within \textit{Cadence}, the results reported below focus on feedback and observations that relate most closely to the DR workflow.

\subsection{Example Use Case}
\label{sec:usecase}

The DR-enabled \textit{Cadence} system was used to analyze a cohort of 1,732 patients to identify risk factors associated with opiate-related abuse or addiction after a hospitalization.  These patients were selected from a larger set of approximately 10,000 longitudinal electronic health records obtained from the University of North Carolina Clinical Data Warehouse for Health \cite{nc_tracs_north_2015}. Inclusion criteria required a pain diagnosis prior to a hospital discharge. The analysis included all recorded medical events (diagnoses and procedures) that occurred between (a) 1 year prior to a patient's pain diagnosis and (b) the time of the patient's hospital discharge. Opiate-related abuse or addiction after the discharge was selected as the outcome measure of interest.

After the initial query, a visualization of the selected cohort shows that after hospitalization, the opiate-related disorder diagnosis rate is 7\%, and that the cohort contains a gender distribution of 55\% female to 45\% male.  To focus the analysis on risk factors for women, a \emph{Gender=Female} filter is applied to create a second, more narrowly-defined cohort. Based on the scatterplot in Figure~\ref{fig:scenario}-a, \emph{nicotine dependence} is identified as a major risk factor and used to further filter the cohort to focus only on patients with that diagnosis.  The initial query and two subsequent filters produce a history of five cohorts: three defined explicitly by the user's actions (all$\rightarrow$women$\rightarrow$women with nicotine dependence) and two defined implicitly as cohorts excluded by the filters (men $\mid$ women without nicotine dependence).

Noticing an increasing distance between the initial query cohort and current focus cohort (women with nicotine dependence), the user changes to the cohort comparison view seen in   Figure~\ref{fig:scenario}-b (R1). The icicle table confirms that, as expected, the filter constraint dimensions (gender and nicotine dependence) have shifted the most from the baseline to the focus cohort. Looking deeper into the plot, however, she discovers that lower respiratory diseases are over represented.  She adds this dimension to the reweight dimension list and examines the subgroups that this would create in Figure~\ref{fig:scenario}-c (R4).  After noticing the low danger scores for this configuration, the user clicks the rebalance button to apply it to the visualization (R2).  Examining the vector plot in Figure~\mbox{\ref{fig:scenario}-e}, the user sees that the reweighting has widespread effects, with small changes in distance to many dimensions.  However, she notices that the largest changes are decreases in distance (i.e., reductions in selection bias) for other important conditions such as COPD.  This observation is confirmed by selecting blue vectors and examining the changes in prevalence rates due to reweighting as shown in Figure~\ref{fig:scenario}-f~(R3).

Returning to the original \textit{Cadence} visualization of clinical event sequences in Figure~\ref{fig:scenario}-g, the user discovers that a key risk factor of heart failure (circled in blue) she had noticed earlier is no longer an exceptional risk factor. She adjusts the rebalance slider to turn the weighting on (Figure~\ref{fig:scenario}-g) and off (Figure~\ref{fig:scenario}-h) to see the impact of the bias correction: heart failure's correlation with outcome drops from well over 0.2 to about 0.18, and it no longer stands out as an outlier with high correlation (R3). As shown in the accompanying video figure, the analyst can then continue exploring alternative reweighting solutions, leveraging the danger score to avoid inappropriate reweighting~(R2,~R4).

\subsection{Domain Expert Interviews}

We provided a hands-on demonstration of the DR-enabled \textit{Cadence} system and conducted semi-structured interviews with medical research experts to gather qualitative feedback on the methods presented in this paper.  This section presents a thematic analysis of the interview findings.

\vspace{-0.05cm}
\subsubsection{Participants and Process}
\vspace{-0.05cm}

Three participants were recruited to take part in the DR demonstrations and semi-structured interviews.  All three were health-focused professionals with advanced degrees (one MD, two PhDs) who regularly work with health data from the UNC Health system. The MD participant works for the health system's analytics division responsible for surfacing data-driven insights for operational purposes and no longer practices clinically.  In contrast, the two PhD-trained participants work on health-related research activities as employees of the University of North Carolina.  All three participants have experience with interactive data analysis in general and with health data analysis in particular.

Each participant took part individually in a one-hour study session with a single study moderator. Sessions took place via Zoom video conference with screen-sharing and live video due to restrictions on face-to-face interactions during the COVID-19 pandemic.  During the session, participants were given a short introduction to DR and the \textit{Cadence} system, provided with a demonstration of key features, and walked through a sample analysis similar to the use case from Section~\ref{sec:usecase}. Participants then took part in a semi-structured interview during which they could also ask questions of the moderator and explore the user interface based on their own curiosity. Finally, participants were asked to complete a written questionnaire at the end of the session.

\subsubsection{Thematic Analysis of Interview Findings}

The expert interviews captured a wide range of feedback, ranging from high-level reactions to thoughts on general usability to specific requests for future features.  Here we provide an overview of key themes that emerged from these interviews.

{\bf Overall Workflow.} 
Overall, the DR workflow was considered ``very useful'' and a ``really a good tool for epidemiologists.''  In fact, one participant's first reaction to seeing DR in action was to applaud. ``This is fantastic,'' she said, ``I wish we had this right now in the clinical data warehouse.''  They remarked that the ``user interface makes sense'' and that the process was ``very straightforward,'' though with caveats outlined in more detail later in this section (see \emph{Usability and Discoverability}).  In addition, the cohort tree was considered critical as a way to anchor the process: ``I really like the cohort tree'' which acts ``like an anchor [and] helps us see what we are comparing.''

{\bf Managing Task and Data Complexity.}
Complexity is a significant challenge in DR. Selecting the right dimensions for reweighting requires navigating a potentially large, complex set of data dimensions.  Users felt that the icicle table effectively summarizes distribution differences in high-dimensional data, with mouseover details being ``very useful'' for knowing where to expand the visualization. The sorting capability helped users ``prioritize'' dimensions for examination by locating the largest differences at the top, which let users quickly ``ignore those less likely to be helpful'' and made them ``more likely to explore'' by helping them focus on the dimensions most relevant for inspection.  

After reweighting, the users commented that the vector plot ``gives more information'' about the complex effects of reweighting beyond the weighted statistics incorporated into visualizations of a standard workflow. These include the somewhat counterintuitive fact that the distributions of some dimensions often became less similar to the baseline. Most importantly, however, participants found the vector plot ``very useful'' for identifying variables that experienced large reductions in distance as a side effect of the specified reweight dimensions. The reweight set visualization was also perceived as quite intuitive and useful for seeing subset sizes across the different cohorts.

{\bf User Agency.} The DR workflow must balance providing users with agency over the bias correction process while also preventing them from applying statistically questionable weighting solutions.  This is accomplished primarily via the danger score and the balance panel.  Participants were perhaps most excited by these capabilities. ``I like this column,'' said one participant in reference to the balance panel, ``super cool'' said another who continued, ``I really like the reweight danger.... I like getting that feedback.''

However, there were some differences of opinion on how the danger score should be used.  One user liked the idea of warning users but also allowing the use of dangerous configurations: ``absolutely'' users should be allowed to proceed despite a high danger score.  Another participant felt that a strict policy forbidding the use of high-danger reweighting solutions would be better to ``prevent people from getting into trouble later.'' This user suggested using novice and expert modes, with strict enforcement only in the former.
There was also a suggestion to make danger scores visible earlier via the split icicle plot to help users avoid examining problematic configurations. In practice this feature could be computationally prohibitive, but it could help users avoid wasting time with dimensions that cannot be effectively corrected.

Relatedly, two of the three participants tended to think of the danger score as a binary property, such that a set of dimensions was either dangerous or safe. This may be in part due to the use of color coding of danger values above a threshold, but it also reflects the binary nature of the user's decision---should they proceed with the correction for these dimensions or change course?
More effective ways to communicate the nuance of the danger score is an important topic for future work.

Finally, the slider in the balance column that controls the amount of reweighting was seen as an important component in support of user agency:
``I like this adjustable slider. You give people control.''  Another participant said that it 
``gives me feedback that the rebalancing has a real effect.''  While the slider enables users to choose intermediate settings, most commonly the slider was used to turn the reweighting on or off as a binary toggle.  The animated response to rebalancing 
enabled users to apply this toggle to ``see where [the data] moves.''

{\bf Usability and Discoverability.}
Participants found the interface complicated but usable. ``There is so much information,'' but it does ``a great job of showing this all on one screen.''  Interviewees recognized that this was a system for sophisticated users who would be trained to use the software. One ``didn't get it at first'' but found it informative after the introduction.  ``Once you explained it, it was obvious'' how things worked, said another. The third participant said, ``after the walkthrough'' it was ``very intuitive.'' This user suggested that in real deployment a video help section would be useful for those who do not get the opportunity to have face-to-face tutorials.

Users appreciated consistent use of indicators across the interface, including those used to distinguish constrained and reweighted dimensions.  However, as all participants were seeing the software for the first time, there were multiple requests for more aggressive use of legends and help resources.  In particular, participants expressed the need for a place to look up symbols or terms used in the interface.

There was also one participant who felt that certain visual elements, particularly the icicle table, may have tried to show too much information at the same time: ``Too many buttons and controls...it just makes it complicated.''  On closer inspection, however, the user felt that these features were all useful, suggesting that it might be valuable to start with a simpler interface which users could optionally enhance by enabling additional capabilities as needed.  

{\bf Potential for Misinterpretation.}
The additional capabilities provided by DR enable the correction of selection bias effects, however this increased complexity may lead to a larger cognitive load for the user. Care must be taken to apply DR appropriately such that other user-driven artifacts are not introduced due to misinterpretation of the selection bias visualizations or inappropriate reweightings. Future work will investigate supporting robust analysis without overloading the user.

\section{Conclusion}

This paper presented dynamic reweighting (DR), a new approach to selection bias mitigation. DR is embodied by a four-step workflow that augments the traditional visualization process to facilitate the correction of selection bias effects during visual analysis.  DR also includes a series of visualization designs and statistical reweighting methods that together enable the creation of bias-corrected visualizations.  Moreover, DR provides analysts with the tools required to interactively guide and assess the bias correction process. 

A prototype implementation of the DR approach was developed within a visual analysis platform for cohort selection from medical data, and a real-world use case
was presented to showcase both (1) how the bias-correction workflow works in practice and (2) the impact of DR on analysis results. In addition, the paper reported a thematic analysis of qualitative feedback gathered through interviews with medical domain experts. The results suggest that bias mitigation techniques would be a welcomed addition to traditional workflows for visual analysis of medical data, and that the DR approach was useful and understandable (if somewhat complicated). Specific areas for future work include: (1) simplifying the DR interface such that certain components are ``opt-in'' to reduce complexity for novice users, (2) investigating more effective ways to communicate a nuanced understanding of the reweight configuration danger score, and (3) developing ways to help guide the user to useful reweight configurations with less trial and error.

\acknowledgments{The research reported in this article was supported in part by a grant from the National Science Foundation (\#1704018).}

\bibliographystyle{abbrv-doi}

\bibliography{vast2020}
\end{document}